\documentclass[aps,pre,twocolumn,amsmath,amssymb,showpacs,amsfonts]{revtex4}
\usepackage{epsfig}
\usepackage{graphicx}
\usepackage{dcolumn}
\usepackage{bm}

\begin{document}

\title{Distribution of Brownian particles in turbulence}

\author{Itzhak Fouxon$^{1}$}
\author{Eugene Mednikov$^2$}
\affiliation{$^1$ Raymond and Beverly Sackler School of Physics and Astronomy,
Tel-Aviv University, Tel-Aviv 69978, Israel}
\affiliation{$^2$ Blavatnik School of Computer Science, Tel Aviv University, Tel Aviv 69978, Israel}
\begin{abstract}

We consider Brownian particles immersed in the fluid which flow is turbulent. We study the limit where the particles' inertia is weak and 
their velocity relaxes fast to the velocity of the flow. The trajectories of the particles in this case have a strange attractor in the physical space, if the particles' diffusion is neglected. Under the latter condition the singular density of the particles was recently described 
completely. The analysis was done for real turbulence and did not involve the flow modeling. Here we take the diffusion into account showing
how it modifies the statistics. The analysis is performed also for real turbulence. Experimentally testable predictions are made. 

\end{abstract}
\pacs{47.55.Kf, 47.10.Fg, 05.45.Df, 47.53.+n} \maketitle

Lately the problem of spatial distribution of inertial particles in turbulent flows enjoyed much attention of the researchers \cite{MaxeyRiley,Maxey,BFF,FFS1,FalkovichPumir,BecGawedzkiHorvai,Collins,BecCenciniHillerbranddelta,Stefano,Cencini,Klyatskin,Olla,MehligWilkinson}. The subject became timely after the breakthrough in the theoretical understanding of the Lagrangian motion of particles in the flow occurred \cite{review}. While the more traditional subject of tracers that follow the flow is quite complete by now
\cite{review}, the understanding of the behavior of inertial particles is still insufficient. Since all real particles have inertia, and the 
effect of inertia was shown to be a non-perturbative phenomenon that often cannot simply be neglected, then the study of the subject is a very 
important practical question. Theoretical advancement was made mainly for the case of small Stokes
number, where the inertia is weak and the particles "almost" follow the flow. Even in this limit of small Stokes numbers, the particles' distribution is highly
non-trivial. Particles' deviations from the surrounding flow accumulate with time, bringing particles to a strange attractor in
space. This attractor is multi-fractal and its statistics was recently described completely in \cite{IF}, see also \cite{Tzhak}. The description 
was made for the real turbulent flow and it did not involve any modeling. Within this model-independent approach, the statistics of turbulence is 
considered as given but not known. The final answers are expressed with the help of statistical characteristics of turbulence. To apply 
such kind of results one should first estimate or measure the corresponding statistical properties of turbulence, and then substitute them 
in the answer for the particles. This approach was first used in \cite{FFS1} where the correlation codimension of the particles' attractor was 
expressed in terms of different time correlation function of the Laplacian of turbulent pressure. It was continued in \cite{IF} where all 
the rest of quantities, besides the correlation codimension were obtained. Since the statistics of
turbulence is largely unknown \cite{Frisch}, then obtaining such results demands understanding of the universal features of the particles' behavior in the flow, that are independent of the details of the flow statistics.

In this Letter we continue the model-independent approach outlined above to include the Brownian motion of the particles. We again consider the 
case of small Stokes numbers, adding the Brownian motion to the underlying equations of the particles' motion. When the Brownian motion, or diffusion, is taken into account, the density of the particles in the steady state becomes an ordinary smooth, random field. We describe the 
statistics of that field and make a number of predictions that are testable experimentally.

The idea that particles' inertia leads to inhomogeneous spatial distribution dates back to the seminal paper by Maxey \cite{Maxey}. It was observed
that, due to the inertia, heavy particles are pushed out of the vortices and hence they will not distribute uniformly in the flow, like the
inertia-less particles. 

Our analysis is based on the recent finding of a universal description for the singular steady state density of the weakly compressible dynamical systems \cite{Tzhak}. The particles' motion, though governed by Newton's law, admits an effective description in terms of a velocity field in space, where inertia is described by a small compressible correction to the incompressible velocity of the background turbulent flow. This correction leads to a small disbalance of trajectories going in and out of space regions, which accumulates over a long time to a big effect. Thus, compressibility is a singular perturbation. The adequate treatment was performed in \cite{Tzhak}. It was shown there that at sufficiently small scales the inertia is always important, however small it is. When diffusion is taken into account, the steady state density is a smooth function. This function is significantly different from a constant only if the diffusion coefficient of the particles is not too large. Hence we will assume that the Prandtl number is large, that is the fluid viscosity is much larger than the diffusivity of the
particles. It will be clear that large Prandtl numbers are necessary if any significant fluctuations of the density are to be
present in the considered case of small Stokes numbers. 

We pass to the analysis. We assume that the density of the particles is not too large, so the hydrodynamic interaction between the particles and their back reaction on the flow is negligible. Then the analysis is effectively the one of the single-particle motion. We consider a small spherical particle with the radius $a$ and the material density $\rho_p$ suspended in the fluid with the density $\rho$ and the kinematic viscosity $\nu$.
The fluid flow $\bm u(t, \bm x)$ is assumed to be incompressible. We first consider the equations that neglect the Brownian motion of the particles. The Newton law governing the evolution of the particle's position $\bm x(t)$ and the particle's velocity $\bm v(t)$ is assumed to have the form
\begin{eqnarray}&&
\frac{d\bm v}{dt}=-\frac{\bm v-\bm u[t, \bm x(t)]}{\tau},
\end{eqnarray}
where $\tau =  (2/9)(\rho_p/\rho)(a^2/\nu)$ is the Stokes time. Thus we assume that all the forces besides the drag can be neglected \cite{Cencini,Maxey}. Above the Kolmogorov scale $\eta$ \cite{Frisch} is assumed to obey $\eta\gg a$. With no loss we set the total volume and the mass equal to one, so the spatial average of the particles' density $n$ obeys $\langle n\rangle=1$. After transients
\begin{eqnarray}&&
\bm v(t)= \int_{-\infty}^t \exp\left[-\frac{t-t'}{\tau}\right]\bm u[t', \bm x(t')]\frac{dt'}{\tau}.
\end{eqnarray}
We assume $\tau$ is much smaller than the smallest time-scale of turbulence, which is the viscous time-scale $t_{\eta}$, so
the Stokes number $St\equiv \tau/t_{\eta}\ll 1$. 
Then we can substitute the derivative in the integrand by its value at $t'=t$ so
$\bm v(t)\approx \bm u-\tau\left[\partial_t\bm u+(\bm u\cdot \nabla)\bm u\right]$.
Thus at $St\ll 1$ the particle's velocity is determined
uniquely by its position $\bm x(t)$ in space
and one can introduce the particle's velocity field $\bm v(t, \bm x)$, see \cite{Maxey}
\begin{eqnarray}&&
\!\!\!\!\!\!\dot {\bm x}(t)=\bm v\left[t, \bm x(t)\right],\ \ \bm v\equiv \bm u-\tau\left[\partial_t\bm u+(\bm u\cdot \nabla)\bm u\right].\label{basic}
\end{eqnarray}
In the zero inertia limit ${\rm St}\to 0$ the particles follow the incompressible mixing flow of turbulence ${\dot {\bm x}}=\bm u\left[t, \bm x(t)\right]$ and in the steady state they are uniformly distributed in space, so their steady state density $n_s$ equals one. This behavior is characteristic of small dye particles. However, at a small but finite $St$, the small correction $\bm v-\bm u$ gives the particles' velocity field a finite compressibility \cite{Maxey}
\begin{eqnarray}&&\!\!\!\!\!\!\!\!\!\!\!\!\!\!
w\equiv \nabla\cdot \bm v=-\tau\nabla\cdot (\bm u\cdot \nabla)\bm u\neq 0.\label{divergence}
\end{eqnarray}
The physical significance of the non-vanishing divergence can be understood by considering the inertial outward motion of heavy particles in a centrifuge, see \cite{Maxey,MaxeyRiley,FFS1,Olla}, and also \cite{IF} for the relation quantifying the anti-correlation between the vorticity and the particles' density. The steady state density $n_s$ of the particles moving according to Eq.~(\ref{basic}) is singular. It is supported on the
strange attractor - a multi-fractal set in space which is asymptoted by the particles' trajectories at large times, see \cite{IF} for the complete
description of that set. The particles' density has log-normal statistics and it is described completely by the pair-correlation function
\begin{eqnarray}&&\!\!\!\!\!\!\!\!\!\!\!\!\!\!
\langle n_s(0)n_s(\bm x)\rangle \approx \left(\frac{\eta}{x}\right)^{2C_{KY}},\ \ x\ll \eta, \label{pair}
\end{eqnarray}
where angular brackets stand for spatial averaging, and $0<C_{KY}\ll 1$ is the Kaplan-Yorke dimension of the attractor \cite{KY}. The latter is 
estimated as $St^2$ times a positive power of the Reynolds number \cite{FFS1,FalkovichPumir,IF}. The divergence of $\langle n_s^2\rangle$ corresponds to the singularity of the density described above. Note that by $C_{KY}\ll 1$ there are no significant fluctuations of the density already at scales $x\ll \eta$, which is inferred by the saturation $\langle  n_s(0)n_s(\bm x)\rangle \approx 1=\langle n_s\rangle^2$ that holds at $x\ll \eta$. The
presence of the separation scale $l\ll \eta$ where there are no significant fluctuations of the density is the most fundamental property of the 
weakly compressible regime, cf. \cite{Tzhak}.

The above results however cannot be applied directly to the real particles for which the Brownian motion is present always. Here we include the Brownian motion into account, studying how it changes the steady state density. Our starting point is the equation of motion
\begin{eqnarray}&&
\!\!\!\!\!\!\!\!\frac{d\bm x}{dt}=\bm v\left[t, \bm x(t)\right]+\bm \xi,\ \ \langle \xi_i(t)\xi_j(t')\rangle=2\kappa\delta_{ij}\delta(t-t'),\label{fund}
\end{eqnarray}
where $\kappa$ is the diffusion coefficient. Below we assume that the Prandtl number $Pr\equiv \nu/\kappa$ is much
larger than one, $Pr\gg 1$. 

It is clear that the diffusion will smoothen the singular steady state density described above
by making the particles diffuse from the spatial regions, where they would be localized otherwise by the combined action of the flow and the inertia.
The particles' density $n$ obeys the Fokker-Planck equation \cite{Risken}
\begin{eqnarray}&&
\partial_t n+\nabla\cdot\left[\bm v n\right]=\kappa\nabla^2 n, \label{FP}
\end{eqnarray}
where the diffusion coefficient stands near the higher-derivative term, suppressing non-smooth realizations of $n$. Further, we note that because  
diffusion only smears the inhomogeneities making the density smoother, then it follows that there are no significant fluctuations of the density at $l\ll \eta$ also
with diffusion. This property is at the center of the consideration below. 

The key tool of our analysis is the Green's function $G(\bm x| t, \bm x')$ that expresses the density $n(0, \bm x)$ in terms of the density 
$n(t, \bm x)$ at $t<0$ according to 
\begin{eqnarray}&&\!\!\!\!\!\!\!\!
n(0, \bm x)\!=\!\int G(\bm x| t, \bm x')n(t, \bm x')d\bm x',\ \ G(\bm x| 0, \bm x')\!=\!\delta(\bm x-\bm x').\nonumber
\end{eqnarray}
We analyze $G(\bm x| t, \bm x')$ as a function of its initial arguments $t$ and $\bm x'$. As a function of $t$ and $\bm x'$ it satisfies the equation
\begin{eqnarray}&&
\partial_t G+\bm v(t, \bm x')\cdot\nabla_{\bm x'} G=-\kappa\nabla_{\bm x'}^2G,\label{Green}
\end{eqnarray}
which is the Hermitian conjugate to Eq.~(\ref{FP}), see \cite{Risken}. We follow the analysis of \cite{BFF}. At $\kappa=0$ the Green's function is supported on the particle's trajectory $\bm q(t, \bm x)$ obeying
\begin{eqnarray}&&
\frac{\partial \bm q(t, \bm x)}{\partial t}=\bm v[t, \bm q(t, \bm x)],\ \ \bm q(0, \bm x)=\bm x.
\end{eqnarray}
One has 
\begin{eqnarray}&&\!\!\!\!\!\!\!\!
G(\bm x| t, \bm x')=\exp\left(-\int_t^0 w\left[t', \bm q(t', \bm x)\right]\right)\delta\left[\bm x'-\bm q(t, \bm x)\right].\nonumber
\end{eqnarray}
Due to the diffusion the particles that are not located at $\bm q(t, \bm x)$ at time $t$ can also arrive at $\bm x$ at time $0$. This is 
summarized by the diffusive smearing of the support of $G(\bm x| t, \bm x')$ near $\bm x'=\bm q(t, \bm x)$. If we introduce 
$G(\bm x| t, \bm x')=\exp\left(-\int_t^0 w\left[t', \bm q(t', \bm x)\right]\right)\Psi[t, \bm x'-\bm q(t, \bm x)]$ then $\Psi$ obeys the equation that follows from Eq.~(\ref{Green}) and reads
\begin{eqnarray}&&
\partial_t \Psi+\left(\bm v\left[t, \bm x'+\bm q(t, \bm x)\right]-\bm v[t, \bm q(t, \bm x)]\right)\cdot\nabla_{\bm x'}
\Psi
\nonumber\\&&
=-\kappa\nabla_{\bm x'}^2\Psi-w\left[t, \bm q(t, \bm x)\right]\Psi
\end{eqnarray}
We note that at not too large $t$, that will be determined below, the dimensions of the support of $\Psi[t, \bm x]$ are much smaller than 
$\eta$, so that one can approximate $\bm v\left[t, \bm x'+\bm q(t, \bm x)\right]-\bm v[t, \bm q(t, \bm x)]\approx \sigma(t)\bm x'$ where 
$\sigma_{ij}(t)\equiv \nabla_j v_i[t, \bm q(t, \bm x)]$ and $tr \sigma=w\left[t, \bm q(t, \bm x)\right]$. Making the Fourier transform in $\bm x'$ we obtain
\begin{eqnarray}&&
\partial_t \Psi-\sigma_{ij}{\bm k}_i\cdot\nabla_j\Psi=\kappa k^2\Psi,\ \ \Psi(t=0)=1.
\end{eqnarray}
The solution to the above equation can be written as
\begin{eqnarray}&&
\Psi(t, \bm k)=\exp\left[-\frac{1}{2}k_i I_{ij}(t)k_j\right],
\end{eqnarray}
where the symmetric matrix $I(t)$ obeys
\begin{eqnarray}&& \!\!\!\!\!\!\!\!
\frac{dI}{dt}=\sigma I+I\sigma^t-2\kappa,\ \ I(t=0)=0. \label{eqtensor}
\end{eqnarray}
The matrix $I(t)$ admits an explicit expression in terms of the Jacobi matrix $W_{ij}(t, \bm x)\equiv \nabla_j q_i(t, \bm x)$ of the backward in time flow $\bm q(t, \bm x)$. This matrix obeys ${\dot W}=\sigma W$, so that
\begin{eqnarray}&& \!\!\!\!\!\!\!\!
I(t)=2\kappa W(t)\int_t^0W^{-1}(t', \bm x)W^{-1, t}(t', \bm x)dt'W^t(t). \nonumber
\end{eqnarray}
It follows that 
\begin{eqnarray}&&
G(\bm x| t, \bm x')=\exp\left(-\int_t^0 w\left[t', \bm q(t', \bm x)\right]\right)
\nonumber\\&&\times
\int \frac{d\bm k}{(2\pi)^d}\exp\left[i\bm k\cdot 
\left(\bm x'-\bm q[t, \bm x]\right)-\frac{1}{2}\bm k I(t)\bm k\right], \label{Greenanswer}
\end{eqnarray}
where $d$ is the dimension of space (in the physically relevant situation $d=2$ or $d=3$). The above answer coincides with the one of \cite{BFF} with a slightly different notation. The equation makes it clear that $I(t)$ is the 
inertia tensor of the cloud of particles localized at $t=0$ at $\bm x=0$, cf. \cite{BF}. Due to the diffusion the cloud's particles spread 
around $\bm q(t, \bm x)$ over
an ellipsoid which axes are determined by the eigenvalues of $\sqrt{I(t)}$. The largest axis $L$ of this ellipsoid grows as $L(t)=l_{dif}
\exp[|\lambda_d t|]$ where we defined the diffusive scale $l_{dif}$ as $l_{dif}\equiv \sqrt{\kappa t_{\eta}}$. This scale arises as follows. 
The cloud of particles "initially" localized at $\bm x$ starts spreading backward in time. The spread is initially completely diffusive as
the particles moving according to Eq.~(\ref{fund}) have initially very close velocities. In particular, in Eq.~(\ref{eqtensor}) at small $t$, the diffusion term $2\kappa$ is the largest and $I_{ij}(t)\approx 2\kappa |t|\delta_{ij}$. However at times $t$ of order $t_{\eta}$ the diffusive term becomes of the order of $\sigma I$ terms. The subsequent growth was described in \cite{BF}. The application to the weakly compressible case considered here gives
that at $t\gg t_{\eta}$ the scale $L(t)$ grows due to the chaotic exponential separation of the trajectories backward in time. This separation occurs at the principal exponent of the flow backward in time that in our case coincides with the absolute value of the $d-$th Lyapunov exponent \cite{Oseledets} of turbulence $\lambda_d$. This is because the flow is weakly compressible and to the leading order the Lyapunov exponents of $\bm v$ coincide with those of $\bm u$. The latter flow is incompressible, so the principal exponent of the flow backward in time is $|\lambda_d|$.

We now can determine the domain of applicability of Eq.~(\ref{Greenanswer}). This equation was derived assuming the linear size of the support 
of $\Psi$ is much smaller than $\eta$. This means $L(t)\ll \eta$, i. e. Eq.~(\ref{Greenanswer}) applies as $|t|\ll |\lambda_d|^{-1}\ln(\eta/l_{dif})$, cf. \cite{BFF}

We are now ready to find all the single point moments of the density. We use the presence of the separation scale $l\ll \eta$ at which there 
are no appreciable fluctuations of the density. Consider 
\begin{eqnarray}&&
n(0, 0)\!=\!\exp\left(-\int_{t_*}^0 w\left[t', \bm q(t', \bm x)\right]\right)
\nonumber\\&&\times
\int \Psi[t_*, \bm x'-\bm q(t_*, \bm x)])n(t_*, \bm x')d\bm x'
\end{eqnarray}
where $t_*=-|\lambda_d|^{-1}\ln(l/l_{dif})$. The function $\Psi$ in the integrand (which has a unit space integral) performs an effective 
averaging of $n(t_*, \bm x)$ over a region with linear size $l$. Since there are no fluctuations at this scale we have that the result of this 
averaging is just the average density, i. e. one. Thus we obtain 
\begin{eqnarray}&&\!\!\!\!\!\!\!\!\!\!\!
n(0, 0)\approx \exp\left(-\int_{-|\lambda_d|^{-1}\ln(l/l_{dif})}^0 w\left[t', \bm q(t', \bm x)\right]\right).
\end{eqnarray}
The above expression is the same expression that was derived in \cite{IF} for the singular density $n_s$ coarse-grained over the scale $l_{dif}$. Using the result for the statistics of the coarse-grained density obtained in \cite{IF}, we conclude
that the single-point statistics of the density is log-normal,
\begin{eqnarray}&&
\langle n^{\rho}\rangle\approx \left(\frac{\eta}{l_{dif}}\right)^{C_{KY}\rho(\rho-1)}.
\end{eqnarray}
We note that the fact that $l_{dif}$ is defined only up to a factor of order one is of no consequence. Any additional factor $C$ of order one would produce one when $l_{dif}$
is substituted by $Cl_{dif}$ in the above equation, due to $C^{C_{KY}}\sim 1$. We also substituted $l$ by $\eta$ since by definition of $l$ at this scale there are no fluctuations and $(\eta/l)^{C_{KY}}\approx 1$. Above it is assumed that $\rho$ is not too large: very high-order moments are determined by the optimal fluctuations for which isotropic compression of the mass occurs, and compressibility is not small in any sense \cite{BecGawedzkiHorvai,BFF}. Finally, we note that since $\eta/l_{dif}\sim Pr^{1/2}$ then the above answer can be rewritten as 
\begin{eqnarray}&&
\langle n^{\rho}\rangle\approx Pr^{C_{KY}\rho(\rho-1)/2},
\end{eqnarray}
where we again use the fact that factors of order unity become irrelevant after being raised to a small power.

Similar consideration of the different point correlation functions of the density reveals that these correlation functions 
are given by the correlation functions of the singular density $n_s$ coarse-grained over the scale $l_{dif}$. The latter 
correlation functions were computed in \cite{IF}, where they were shown to obey the log-normal statistics. For example the expression for 
the pair correlation function is modified from Eq.~(\ref{pair}) to
\begin{eqnarray}&&\!\!\!\!\!\!\!\!\!\!\!\!\!\!
\langle n(0)n(\bm x)\rangle \approx \left(\frac{\eta}{x+l_{dif}}\right)^{2C_{KY}},\ \ x\ll \eta.
\end{eqnarray}
Though the above formula is obtained by interpolation from $x\ll l_{dif}$ to $x\gg l_{dif}$, it gives an answer which is correct 
approximately. This is again because the correlation function changes insignificantly when $x$ changes by the order of magnitude
due to $C_{KY}\ll 1$. This finishes the consideration of the steady state statistics of the density of particles moving according to Eq.~(\ref{fund}). We note that our analysis that assumed that there are no significant fluctuations of density already at the scale 
$l\ll \eta$ is self-consistent as it is clear from the obtained formulas.

We showed that it is possible to describe the statistics of the density of the particles in real turbulence taking into 
account the Brownian motion of the particles. The result is quite expectable. The statistics of the density is roughly 
the same as the statistics of the coarse-grained density of the particles that do not perform the Brownian motion, provided 
the scale of coarse-graining is chosen to be equal to the diffusive scale $l_{dif}$. Still several non-trivial points must be 
stressed. First and foremost, the Brownian motion is always present for the real particles and clear understanding of its 
effects is necessary. As an application let us derive the criterium for the existence of significant fluctuations of the density. 
The latter exist only if $\langle n^2\rangle$ is significantly larger than
one, i. e. $C_{KY}\ln Pr \gtrsim 1$. We observe that the dependence on the Prandtl number is only logarithmic. Thus since $C_{KY}$ is 
proportional to $St^2$ times a power of the Reynolds number, then the latter must be substantial to have significant fluctuations 
at $St\ll 1$, cf. \cite{FalkovichPumir}. 

The second reason why the analysis is important is methodological. We saw that a different method was used to derive 
the statistics than in the previous works. Since the statistics of the distribution of particles with no Brownian motion is an idealization, the singular
distribution obtained in that limit must be considered as the $\kappa\to 0$ limit of the distribution obtained here. In other 
words, the physically sound way of thinking of the singular measure $n_s$ supported on the strange attractor of the particles in space 
is as the limit
\begin{eqnarray}&&
n_s=\lim_{\kappa\to 0}n(\kappa),
\end{eqnarray}
where $n(\kappa)$ is the smooth steady-state density obtained in this work, where $\kappa$ is considered as a parameter. 

Our analysis can be generalized directly to the case of light particles as in \cite{IF}.
A possible future application of the methods developed here is to the study of the distribution of inertial particles at $St\sim 1$, 
and it is the subject for the future work.   

This work was supported by the BSF Grant No. 2010314. The research of E. M. was supported in part by an ERC Advanced grant.

\end{document}